\documentclass[aps,pre,showpacs, twocolumn]{revtex4-1}
\usepackage{graphicx}
\usepackage{amsmath}
\usepackage{amssymb}
\usepackage{tabu}
\usepackage{color}
\usepackage{hyperref}
\usepackage{multirow}
\usepackage{setspace}

\hypersetup{pdftitle={Universality classes of quantum chaotic dissipative systems},
           pdfauthor={Ambuja Bhushan Jaiswal, Ravi Prakash, Akhilesh Pandey},
           colorlinks=true,
           citecolor=blue,
           urlcolor=blue}
\usepackage{booktabs} 

\graphicspath{{./images/}}
\begin{document}
\title{Universality classes of quantum chaotic dissipative systems}
\author{Ambuja Bhushan Jaiswal} \email{abjaiswal.jmi@gmail.com}
\author{Ravi Prakash} \email{raviprakash.sps@gmail.com}
\author{Akhilesh Pandey}\email{ap0700@mail.jnu.ac.in}\email{apandey2006@gmail.com}
\affiliation{School of Physical Sciences, Jawaharlal Nehru University, New Delhi -- 110067, India}
\begin{abstract}
We study the ensemble of complex symmetric matrices. The ensemble is useful in the study of effect of dissipation on systems with time reversal invariance. We consider the nearest neighbor spacing distribution and spacing ratio to investigate the fluctuation statistics and show that these statistics are similar to that of dissipative chaotic systems with time reversal invariance. We show that, unlike cubic repulsion in eigenvalues of Ginibre matrices, these ensemble exhibits a weaker repulsion. The nearest neighbor spacing distribution exhibits $P(s) \propto -s^3 \log s$ for small spacings. We verify our results for quantum kicked rotor with time reversal invariance. We show that the rotor exhibits similar spacing distribution in dissipative regime. We also discuss a random matrix model for transition from time reversal invariant to broken case. 
\end{abstract}


\maketitle
\section{Introduction}
The quantum mechanical behavior of dissipative quantum systems are of great interest \cite{rivas, breuer, davies}. For quantum chaotic systems, ensembles  of asymmetric complex matrices (the Ginibre matrices) are helpful to study the effect of dissipation on statistical properties. We will consider the random symmetric complex matrices and their application in the study of effect of dissipation in quantum chaotic systems with time reversal invariance (TRI).

There has been a lot of work on hermitian and unitary random matrices \cite{rmp-apandey, bgs-2, rmp-beenakker, guhr,stockmann, mlmehta, oxford}. The various ensembles of hermitian matrices viz. Gaussian Orthogonal Ensembles (GOE), Gaussian Unitary Ensemble (GUE), and Gaussian Symplectic Ensemble (GSE) give real eigenvalues and are applicable in the study of the Hamiltonians of conservative quantum chaotic systems. GUE is applicable when TRI is broken. GOE is applicable when TRI and rotational symmetry are both preserved. When TRI is preserved but rotational symmetry is broken, GOE and GSE are applicable for system with integral and half-integral spins respectively. Similar classification applies to the ensemble of unitary matrices viz. Circular Orthogonal Ensembles (COE), Circular Unitary Ensemble (CUE), and Circular Symplectic Ensemble (CSE). They are used in the study of the evolution operators for quantum chaotic maps which arise from time-periodic Hamiltonians. System of Quantum kicked rotor (QKR) provides a nice demonstration of COE and CUE \cite{izrailev} The above three classes of ensembles of both types are invariant under orthogonal, unitary, and symplectic transformations respectively. Moreover, in both types, the matrices are symmetric, asymmetric, and quaternion self dual respectively. These are characterized by the Dyson  parameter $\beta$  with values $1,2,$ and $4$ respectively. These ensembles provides universal eigenvalues fluctuation statistics. For example, the nearest neighbor spacing distribution (nnsd), viz., the distribution of consecutive eigenvalues and eigenangles for both types of ensembles show  Wigner distribution with linear, quadratic, and quartic level repulsions for the three $\beta$ classes respectively. In contrast quantum integrable systems show Poisson statistics where level clustering is observed \cite{mlmehta, oxford, bgs, berry1977}.The Poisson distribution may be interpreted as the  $\beta = 0$ case.

Quantum chaotic  dissipative systems (QCDS) are studied in the framework of Ginibre ensembles (GinE) \cite{ginibre, oxford, edelman-1997, Forrester-2007}. These ensembles do not follow any Hermiticity or unitarity, but consist of  matrices with general complex elements. Eigenvalues for such ensembles lie in the complex plane \cite{ginibre, mlmehta, lehmann-1991}. The imaginary part of the eigenvalues and the eigenangles are considered as a manifestation of dissipation in the system. The spacing distribution for the Ginibre ensemble shows cubic repulsion in the eigenvalues \cite{haakebook} and is  verified in dissipative quantum kicked rotor (DQKR) without TRI \cite{qdissipation,ravi-2015,qdissipation2,haakebook, d_braun}.

In this paper, we consider the fluctuation statistics of DQKR when TRI is preserved. The  quantum kicked rotor (QKR) with TRI preserved and TRI broken are modeled by  COE and CUE respectively \cite{izrailev, pramana-kicked-rotor, pragya-cue}. In a similar way, we introduce the ensemble of symmetric Ginibre matrices (symm-GinE) as a random matrix model to study the TRI case of DQKR. We will show that the fluctuation statistics obtained in DQKR is different from the TRI breaking case.

\section{Four classes of complex random matrices} \label{sec-rm-4-classes}
Analogous to the above four cases of the circular and hermitian random matrix ensembles, we have four classes for dissipative systems. Analogous to the Poisson statistics is the distribution of uncorrelated complex numbers. The corresponding nnsd exhibits the Wigner distribution with linear repulsion \cite{haakebook}. The dissipative systems with no time reversal invariance are represented by complex asymmetric matrices (the Ginibre ensemble) and the corresponding nnsd exhibit universal cubic repulsion. We will show that effect of dissipation on time reversal invariant systems can be studied with ensemble of complex symmetric matrices. We also believe that the ensemble of complex quaternion self dual matrices will be applicable in the study of dissipation in time reversal invariant systems. The difference between these two is decided by rotational symmetries in the above Gaussian and Circular ensembles. We again represent the four classes by the parameter $\beta$. The parameter has the value $\beta = 0$ for complex diagonal matrices, $\beta = 1$ for complex symmetric matrices, $\beta = 2$ for general complex matrices (the Ginibre matrices) and $\beta = 4$ for the self dual complex quaternion matrices. 

We consider ensembles of $N$-dimensional matrices $M$ with elements distributed as  complex Gaussian variables of zero mean and variances $v^2$ for both real and imaginary parts. The joint probability distribution (jpd) of these matrices can be written as: 
\begin{align}
\label{eq-jpd-mat-2dim}
\nonumber P(M) & \propto \exp\left[-\frac{1}{2v^2}\text(\text{Tr}~M^\dag M)\right] \\
& = \exp \left[-\frac{1}{2v^2} \sum_{j,k} |M_{j,k}| ^2 \right],
\end{align}
where the $M_{jk}$, $(j,k = 1,...,N) $ are complex numbers for $\beta = 0, 1,2$ and complex quaternion numbers for $\beta = 4$. For $\beta = 0$ , we have $M_{i,j} = M_{j,i} = 0$ , for $\beta = 1$, we have $M_{i,j} = M_{j,i}$ , and for $\beta = 2$ , $M_{i,j}$ and $M_{j,i}$ are independent.  For $\beta = 4$ , $M_{j,k}$ are the quaternions with the property $M_{jk}^D = M_{kj}$, where $D$ represents the dual of the quaternion. A quaternion number $q$ is written as $q = q_{0}e_{0}+ q_{1}e_{1} + q_{2}e_{2} + q_{3}e_{3}$ where $e_{0}, e_{1},e_{2},e_{3}$ are the quaternion units. The dual of $q$ is given by $q^D = q_{0}e_{0}- q_{1}e_{1} - q_{2}e_{2} - q_{3}e_{3}$ and complex conjugate of $q$ is given by $ q^* = q_{0}^*e_{0}+ q_{1}^*e_{1} + q_{2}^*e_{2} + q_{3}^*e_{3}$. In the matrix representation, the quaternions are replaced by their 2-dimensional matrices \cite{mlmehta}.

\section{Two dimensional Random Matrices} \label{sec-2-dim-rm}
We first consider the spacing distribution for various universality classes in two-dimensional complex random matrices $(N = 2)$. The spacing $s = |z_1 - z_2|$ of the eigenvalues $z_1$ and $z_2$ can be written in terms of the matrix elements as
\begin{align}
\label{eq-sp-2dim}
s_M = \sqrt{\left| [(M_{11} - M_{22})^2 - 4M_{12}M_{21} \right ]|}.
\end{align}
The spacing distribution $p(s,\beta)$  for $\beta$  is given by,
\begin{align}
\label{eq-sp-beta0-2dim}
p(s,\beta) & \propto \int\ldots\int\delta(s-s_M) P(M)~dM,
\end{align}
with $v^2$ chosen such that the average spacing is unity.

\begin{figure}[!h]
\includegraphics[width=0.76\linewidth, clip]{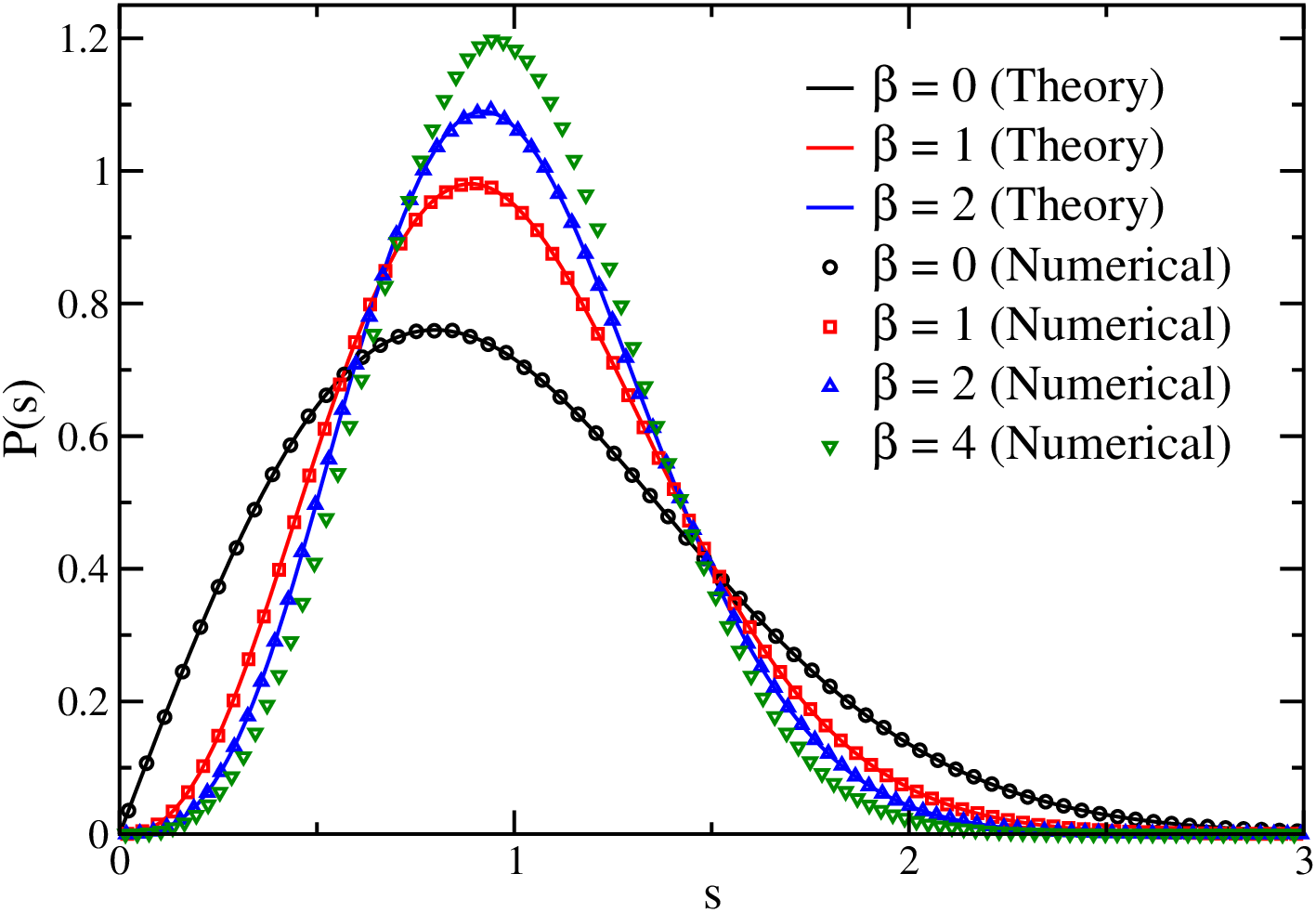}
\caption{Nearest neighbor spacing distribution for two-dimensional matrices. Theory is from (\ref{eq-sp-beta0-2dim-2},\ref{eq-sp-beta1-2dim-1},\ref{eq-sp-beta2-2dim}) and numerical results are from simulation of two-dimensional matrices.\label{fig-sp-beta}}
\end{figure}

$s_{M}$ can be written as $\left| \sum w_j^2\right|^{1/2}$ with $(j=1,...,\beta+1)$, where $w_j$ are independent Gaussian variables with variances $v^2$ for both real and imaginary parts. Thus $p(s, \beta)$ can be written as:

\begin{align}
\label{eq-sp-beta-2dim}
\begin{split}
p(s, \beta) & \propto \int \ldots \int \exp\left (-\frac{1}{2v^2}\sum_{j = 1}^{\beta + 1} \left| w_j\right|^2 \right) \\
 &\quad\times\delta \left(s-\left|\sum_{j=1}^{\beta+1}  w_j^2\right| ^{1/2}\right) \prod_{j=1}^{\beta+1}d^2w_j.
 \end{split}
\end{align}
The compact expression for the spacing distribution can be derived for $\beta = 0,1,2$ from (\ref{eq-sp-beta-2dim}). For $\beta = 0$ case, we get the Wigner distribution
\begin{align}
\label{eq-sp-beta0-2dim-2}
p(s,0) = \frac{\pi}{2} s \exp(-\pi/4 s^2).
\end{align}
For $\beta = 1$ case, we obtain
\begin{align}
\label{eq-sp-beta1-2dim-1}
P(s,1) & = c_1 s^3 K_0\left(c_2 s^2\right),
\end{align}
with
\begin{align}
c_1 = \frac{1}{2^{13}} \left[\Gamma\left( \frac{1}{4}\right)\right]^8; c_2 = 2 \left[\Gamma\left( \frac{5}{4}\right)\right]^4.
\end{align}
Here $K_0(s)$ is the zeroth order modified Bessel function of the second kind \cite{bessel-function}
\begin{align}
\label{eq-k0}
K_0(s) = \int_s^\infty \frac{1}{\sqrt{x^2 - s^2}} e^{-x} dx.
\end{align}
For $\beta = 2$ case we have \cite{haakebook},
\begin{align}
\label{eq-sp-beta2-2dim}
p(s,2) = 2 \left(\frac{9\pi}{16}\right)^2 s^3 \exp \left(- \frac{9\pi}{16} s^2 \right).
\end{align}
 We have scaled the variance in all four cases so as to get normalized spacing distribution with mean spacing one. For small spacing we see from (\ref{eq-sp-beta1-2dim-1},\ref{eq-sp-beta2-2dim}) that Ginibre ensemble exhibits cubic repulsion, $P(s) \propto s^3$ whereas ensemble of complex symmetric matrices follows $P(s) \propto \left(-s^3 \log(s)\right)$ \cite{bessel-function}. The nearest neighbor spacing distribution for all four classes are shown in Fig.~\ref{fig-sp-beta}.

Unlike the Gaussian ensemble for conservative systems, the spacing distribution for large dimension matrices are not similar to that of two-dimensional case except for small spacings, but exhibit universality in their respective classes. The $\beta = 0$ case however remains the same in large dimensional matrices.

\section{Ginibre Ensemble for large dimensions - Brief review} \label{sec-Ginibre}
Ginibre ensemble consist of asymmetric matrices with complex entries. The matrix elements follow the Gaussian distribution. The jpd for the Ginibre matrices is given by (\ref{eq-jpd-mat-2dim}) with $v^2 = 1/2$:
\begin{equation}
\label{eq-jpd-mat-beta2}
P(M) \propto \exp(-\text{Tr}~M^\dag M).
\end{equation}
The eigenvalues of such matrices lie in the complex plane. In order to obtain eigenvalue jpd, the matrix is transformed to eigenvalue-eigenvector space followed by the integration over eigenvector variables. The eigenvalue jpd for Ginibre ensemble is given by \cite{mlmehta, haakebook, ginibre},
\begin{equation}
\label{eq-jpd-ev-beta2}
P(z_1, z_2, \ldots, z_N) = C \prod_{1\leq j < k \leq N} \left | z_j - z_k  \right |^2 e^{-\sum_{l=1}^N |z_l|^2}.
\end{equation}
For Ginibre ensemble, the spectral density is constant for large values of $N$ and is given by,
\begin{equation}
\label{eq-r1-beta2}
R_1(z) = \left \{
\begin{array}{ll}
1/\pi, & |z| \leq \sqrt{N}, \\
0, & |z| > \sqrt{N}.
\end{array}
\right .
\end{equation}

The spacing distribution can also be evaluated from jpd of eigenvalues. One defines $p(s,\beta)$ to represent the spacing distribution of nearest neighbor distance for each particle in the complex plane i.e. for each eigenvalues $z_0$ one can find an eigenvalue $z_1$ for which $s = |z_0 - z_1|$ is minimum. For large N the nnsd for the Ginibre ensemble is given by \cite{haakebook},
\begin{equation}
\label{eq-sp-beta2}
p_N(s,2) = - \frac{\mathrm d}{\mathrm ds} \prod_{n=1}^\infty \left[ e_n(s^2) e^{-s^2}\right],
\end{equation}
where
\begin{equation}
e_n(x) = \sum_{k=0}^n \frac{x^k}{k!}.
\end{equation}
For small spacings, $p_N(s,2)$ can be written as,
\begin{equation}
\label{eq-sp-beta2-1}
p_N(s,2) = 2s^3 - s^5 + \frac{1}{3}s^7 - \frac{11}{12}s^9 + \ldots.
\end{equation}

Thus the nearest neighbor spacing distribution exhibits cubic repulsion for small spacings. 

\section{Numerical Results for large N} \label{sec-symm-Gin}
 For numerical results we use (\ref{eq-jpd-mat-beta2}) for all three $\beta$ and consider ensembles of $10000$ matrices with $N = 500$. For $\beta=1$, we consider complex symmetric matrices with real and imaginary  entries of the off-diagonal matrix elements are independently distributed as Gaussian variables with mean $0$ and variance $1/2$. In this case the diagonal matrix elements have variance twice that of the off-diagonal elements. For $\beta = 2$, every element is a complex Gaussian variable with mean and variance $1/2$. For $\beta = 4$, we need one symmetric and three anti-symmetric complex matrices with the same mean and variance as above. 
 
\begin{figure}[!h]
  \includegraphics[width=0.9\linewidth, clip]{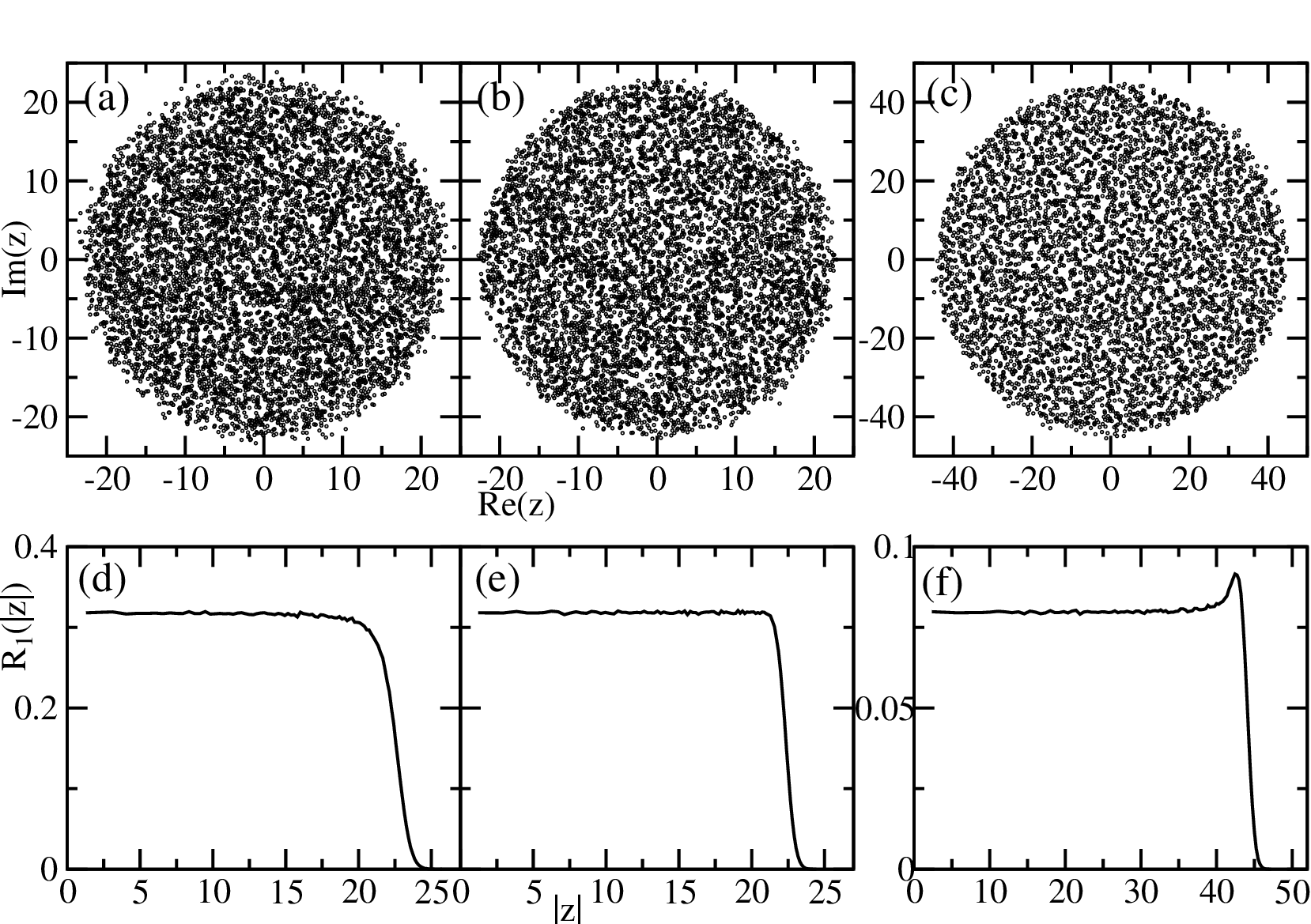}
  \caption{Eigenvalue scatter plot for the eigenvalues of (a) Complex symmetric matrices ($\beta = 1$), (b) the Ginibre matrices ($\beta = 2$) and (c) the self-dual matrices of complex quaternions ($\beta = 4$). Their density profiles are shown in (d), (e) and (f) respectively.\label{fig-ev-r1-cs-a0-N}}
 \end{figure}

 We diagonalize the matrices using standard LAPACK routines \cite{lapack}. The eigenvalues are uniformly distributed in circle of radius $\sqrt{N}$ for both complex symmetric and asymmetric (Ginibre) matrices. For $\beta = 4$, there are $N$-distinct eigenvalues, each doubly degenerate, and are distributed uniformly in a circle of radius $2\sqrt{N}$. In Fig.~\ref{fig-ev-r1-cs-a0-N}, we show the eigenvalues scatter plot for $\beta = 1,2$. The eigenvalue distribution is isotropic. We also plot the radial density $R_1(|z|)$, normalized to $N$ ({\it i.e.,} $\int_0^\infty 2\pi r R_1(r)~dr = N$), in the same figure. 
 
We evaluate the nearest neighbor spacing distribution for both the systems. The solid lines in Fig.~\ref{fig-sp-kr-cs-beta12} show the nearest neighbor and next nearest neighbor spacing distributions for $\beta = 1, 2$ and $4$. It may be noted that this letter is concerned with $\beta = 1,2$ and we show the spacing distribution for $\beta = 4$ for completeness. There exist systems where spectral density may not be uniform \cite{ravi-2015} and one require unfolding method to remove the global variations. In order to unfold the spectra of such cases, we scale the each spacing, $s$, by $\sqrt{\pi R_1}$ to get the spectral density similar to that of the Ginibre ensemble, where $R_1$ is the average spectral density around the eigenvalue pair. We will deal with non-uniform density in quantum kicked rotor discussed ahead.

\section{Quantum kicked rotor} \label{sec-qkr}
The Hamiltonian for kicked rotor is defined as,
\begin{equation}
\label{eq-hamiltonian-qkr}
H = \frac{(p+\gamma)^2}{2m} + \kappa \cos(\theta + \theta_0) \sum_{n = -\infty}^\infty \delta(t - n),
\end{equation}
where $\gamma$ and $\theta_0$ are time reversal and parity breaking parameters. For sufficiently large values of kicking parameter, $\kappa (\gtrsim 10)$, the classical kicked rotor exhibits chaotic motion.
The quantum mechanical analogue of classical chaotic motion can be studied using the time evolution operator, given by, $U = BG$, where $B = \exp\left[-i\left( \kappa \cos(\theta + \theta_0)\right)/\hbar\right]$ and $G = \exp\left[-i(p + \gamma)^2/(2\hbar)\right]$ with $\theta,p$ the position and momentum operators respectively. For values of $\kappa^2/N$ of the order $O(10)$, classical system becomes chaotic but quantum system shows Poisson statics because of localization effect \cite{pramana-kicked-rotor,pragya-cue, Fyodorov-1991, Casati-1990}. For sufficiently large values of $\kappa^2/N$ i.e., $O(1000)$, the corresponding quantum system displays chaos and follows the circular ensemble models \cite{izrailev,pramana-kicked-rotor,pragya-cue}.

We apply torus boundary conditions so that both $\theta$ and $p$ are discrete. We set $\hbar = 1$ and consider $N$-dimensional model. In the position representation, the $B$ operator is given by
\begin{equation}
\label{eq-B-qkr}
B_{mn} = \exp\left[-i \kappa \cos\left(\frac{2\pi m}{N} + \theta_0\right)\right] \delta_{mn},
\end{equation}
and the $G$ operator is given by
\begin{equation}
\label{eq-G-qkr}
G_{mn} = \frac{1}{N}\sum_{l=-N^\prime}^{N^\prime} \exp\left[-i\left(\frac{l^2}{2} - \gamma l + 2\pi l \left(\frac{m-n}{N}\right)\right)\right].
\end{equation}
Here $N^\prime = (N-1)/2$ and $m,n = -N^\prime, -N^\prime+1,\ldots,N^\prime$. Thus the evolution operator can be written in position basis as \cite{izrailev},
\begin{align}
\label{eq-floquet-qkr}
\nonumber U_{mn} &= \frac{1}{N} \exp\left[-i \kappa \cos\left(\frac{2\pi m}{N} + \theta_0\right)\right] \\
 \times & \sum_{l=-N^\prime}^{N^\prime} \exp\left[-i\left(\frac{l^2}{2} - \gamma l + 2\pi l \left(\frac{m-n}{N}\right)\right)\right].
\end{align}
The above evolution operator is unitary. In chaotic regime, the nnsd for this operator is similar to that of COE (CUE) for $\gamma = 0$ $(0 \ll \gamma < 1)$.

\section{Dissipative quantum kicked rotor} \label{sec-dqkr}
We introduced a dissipation term in the quantum kicked rotor. The dissipation operator, $D$, is given by,
$D(\alpha) = e^{-\alpha p^2}$, where $\alpha$ is a control parameter for dissipation. The evolution operator for dissipative kicked rotor can be written as, $U(\alpha) = BGD$ and corresponding matrix elements for the Floquet operator in position basis can be written as,
\begin{align}
\label{eq-floquet-dqkr}
\nonumber&  F_{mn}(\alpha) = \frac{1}{N} \exp\left[-i\kappa \cos\left(\frac{2\pi m}{N} + \theta_0\right)\right]\\
& \times  \sum_{l=-N^\prime}^{N^\prime} \exp\left[-i\left(\frac{1 - i \alpha}{2}l^2 - \gamma l + 2\pi l \left(\frac{m-n}{N}\right)\right)\right].
\end{align}
The Floquet operator is no longer unitary. The eigenvalues starts falling towards center and constitute a ring like structure. We have studied the time reversal broken ($\gamma \neq 0$) case for this system in \cite{ravi-2015}.

\begin{figure}
\includegraphics[width=0.8\linewidth, clip]{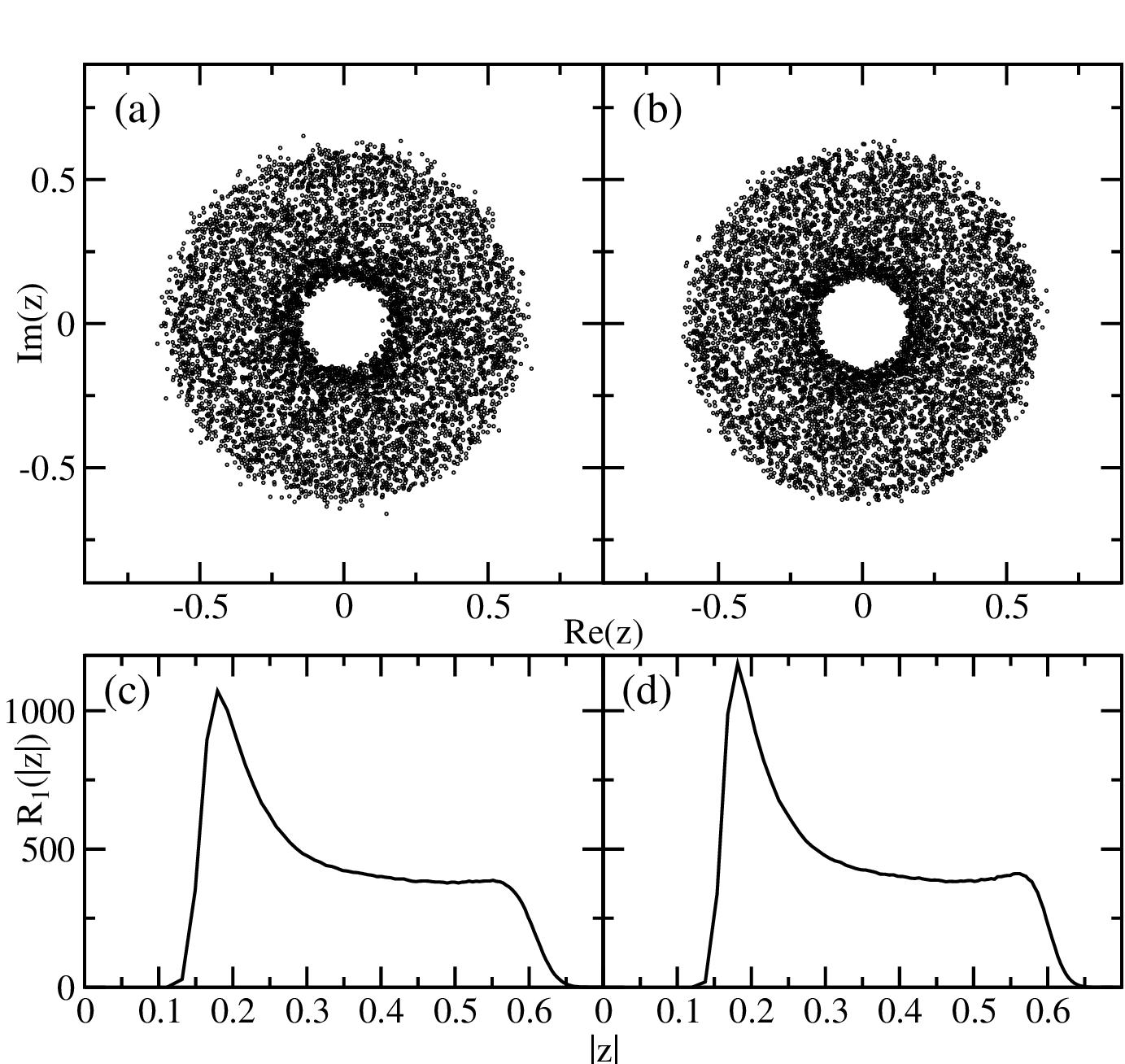}
\caption{Scatter plot for the eigenvalues of the Floquet opertor for (a) Time reversal invariant ($\gamma = 0$) case and (b) Time reversal non-invariant ($\gamma = 0.7$) case. The density profiles for both the cases are shown in (c) and (d) respectively.\label{fig-ev-r1-kr-g0-0.7}}
\end{figure}

\begin{figure}[!h]
\includegraphics[width=0.8\linewidth, clip]{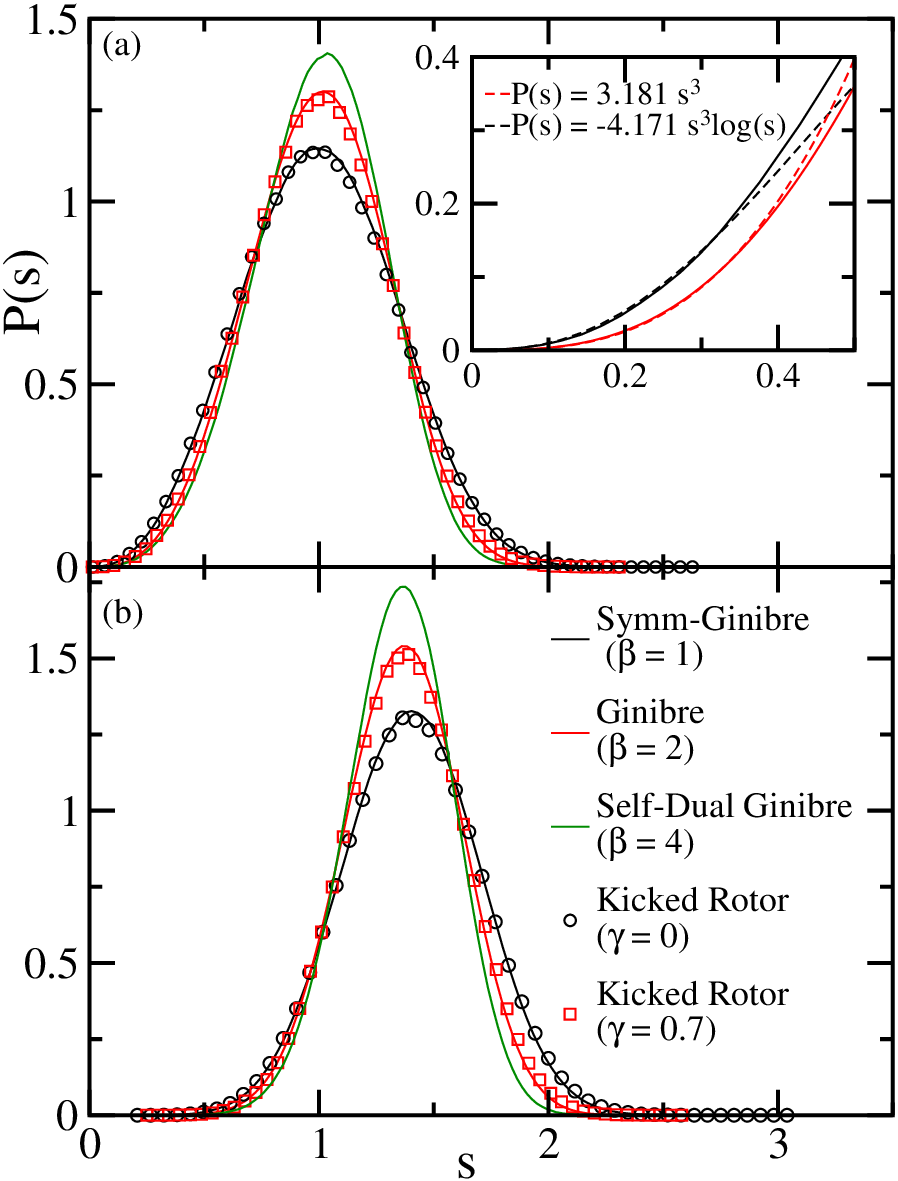}
\caption{(a) The nearest neighbor and (b) next nearest neighbor spacing distribution for kicked rotor with $\gamma = 0$ (circle) and $0.7$ (square) and its agreement with that of symm-Ginibre and Ginibre matrices (solid lines) respectively. Dashed lines in inset of (a) show best fit curve for $\beta = 1$ and $2$ cases.\label{fig-sp-kr-cs-beta12}}
\end{figure}

The time reversal preserved case corresponds to $\gamma = 0$. We construct the spectra using (\ref{eq-floquet-dqkr}) with $\gamma = 0$ and $N = 501$. The spectral density is not uniform in this case as shown in Fig.~\ref{fig-ev-r1-kr-g0-0.7}.
To avoid the errors in unfolding due to non-uniform density, we consider nearly uniform part of the spectra, viz. the spectra in a ring of inner and outer radius 0.255 and 0.520 respectively, i.e., considering approximately $50\%$ eigenvalues of spectra. We thus calculate the nearest neighbor spacing distribution. The nnsd and next nnsd are in an excellent agreement with the spacing distribution obtained from complex symmetric matrices as shown in Fig.~\ref{fig-sp-kr-cs-beta12}. The spacing distribution for dissipative quantum kicked rotor (DQKR) with time reversal broken ($\gamma = 0.7$) and its agreement with the Ginibre ensemble is also shown in the same figure for completeness.

\section{Ratio Test} \label{sec-ratio-test}
In the case where eigenvalues lie on the real line or circle, the spacing distribution of the ensembles can be computed relatively easy. This is due to the unfolding procedure which works quite well in one dimensional spectra. In case of Ginibre ensemble and Symmetric-Ginibre ensemble the spectra we obtained is two-dimensional. Due to the limitations of unfolding procedure we are constrained to use a short range of spectra of the ensemble. In order to use a large range of spectra to study the distribution we take the ratios of the spacings, and evaluate spacing ratio in two ways. In the first case we take the ratio of nearest and next nearest spacing of the spectra and call it type - I ratio. In the second case we take the ratio of spacing of nearest neighbor of a spectra and the spacing of the nearest neighbor from the said nearest neighbor and call it type – II ratio. In both the cases we consider the spectra in a ring of inner and outer radius $0.203$ and $0.601$ respectively, i.e., considering about $87 \%$ eigenvalues of spectra. The average ($m$) and variance ($\sigma$) of the ratio we obtained is shown in the Table \ref{table-ratio-test}. We again see an excellent agreement of spacing ration for quantum kicked rotor with that of random matrix ensemble for both TRI preserved (correspond to $\beta = 1$) and TRI broken (correspond to $\beta = 2$) cases.
\begin{table}[!h]
\caption{Comparison of mean and variance of ratio of spacings.\label{table-ratio-test}}
\begin{ruledtabular}
 \begin{tabular}  {r c c c c}
 & \multicolumn{2}{c}{Type-1} & \multicolumn{2}{c}{Type-2}\\
&  $m$ & $\sigma\times10^{2}$ & $m$ & $\sigma\times10^{2}$ \\
\hline
\textbf{DQKR $\gamma = 0.0$} &  $0.7232$ & $3.8789$ & $0.8995$ & $3.1567$ \\
\textbf{RMT $\alpha = 0.0$}  & $0.7213$ & $3.9046$ & $0.8990$ & $3.1701$ \\
\hline
\textbf{DQKR $\gamma = 0.7$}    & $0.7397$ & $3.5108$ & $0.9084$ & $2.6903$ \\
\textbf{RMT $\alpha = 1.0$}  & $0.7371$ & $3.5415$ & $0.9086$ & $2.6857$
\end{tabular}
\end{ruledtabular}
\end{table}

\section{Intermediate ensembles and their relation with dissipative quantum kicked rotor} \label{sec-crossover}
\begin{figure}[!h]
\includegraphics[width=\linewidth, clip]{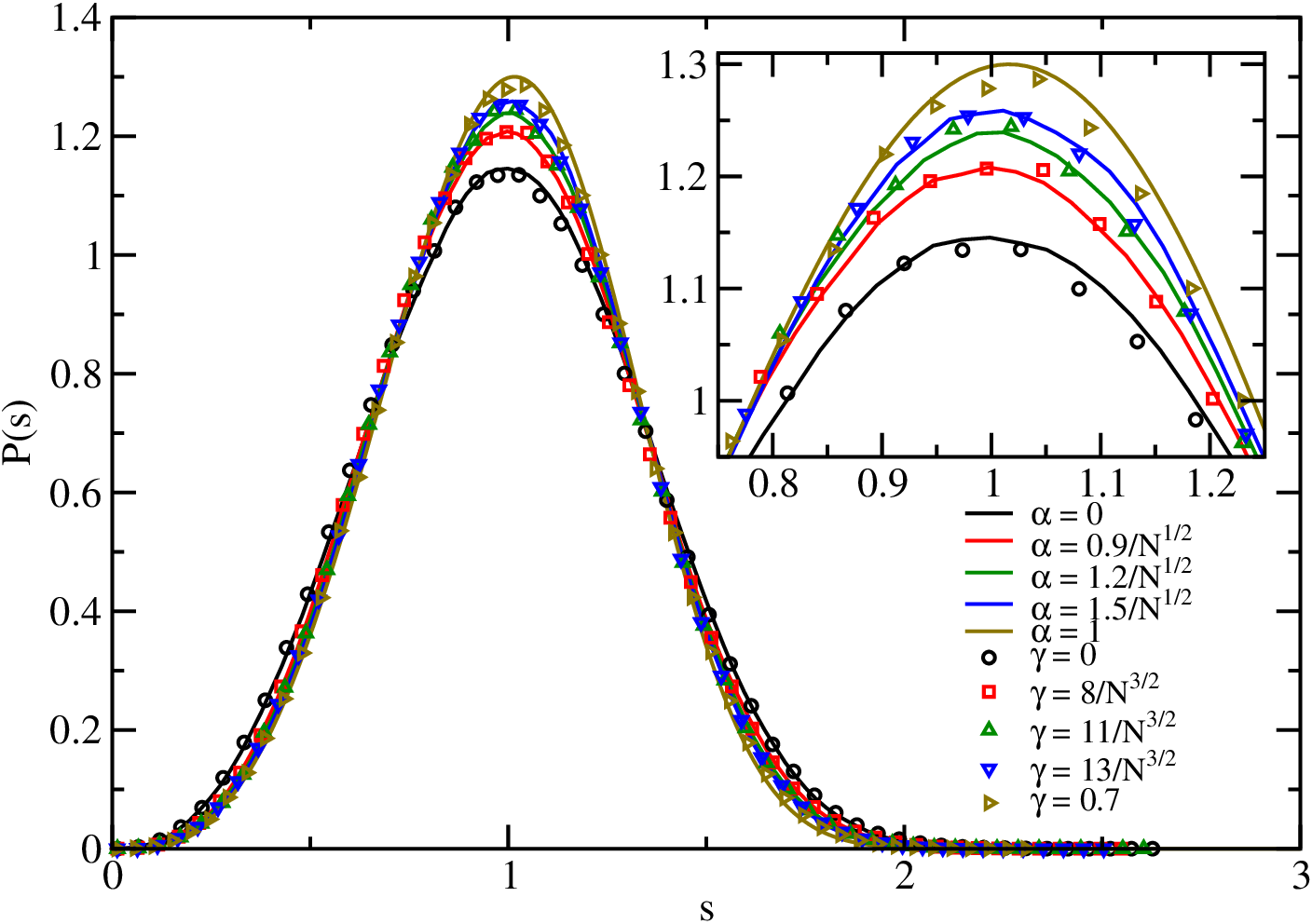}
\caption{Nearest neighbor spacing distribution for intermediate cases and their agreement with random matrix models. \label{fig-sp-kr-cs-beta-intermediate}}
\end{figure}

The intermediate cases of kicked rotor with time reversal invariance weakly broken can be modeled with the linear combination of symmetric and antisymmetric matrices which act as a crossover between symmetric and the Ginibre ensemble. The intermediate matrices $M$ can be defined as
\begin{equation}
\label{eq-rmm-model}
M = \frac{1}{\sqrt{1 + \alpha^2}} S + \frac{\alpha}{\sqrt{1 + \alpha^2}} A,
\end{equation}
where $S$ and $A$ are complex symmetric and complex anti-symmetric matrices. For $\alpha = 0$, we get complex symmetric matrices and for $\alpha = 1$ we get Ginibre matrices. Note that variance of distribution for elements of matrix $M$ is independent of $\alpha$. We shown in Fig.~\ref{fig-sp-kr-cs-beta-intermediate} the spacing distribution for DQKR with various values of TRI breaking parameter $\gamma$ and also find the corresponding best suitable value for crossover parameter $\alpha$.
For a quantitative analysis we show here the mean and variance of different plots in the Table \ref{table-sp-compare}. Here $m$ and $\sigma$ represents the mean and variance of the spacing distribution. The subscripts $0,1$ represent the nnsd and next nnsd respectively.
\begin{table}[!h]
\caption{Comparison of mean and variance for several spacing distributions. \label{table-sp-compare}}
\begin{ruledtabular}
 \begin{tabular}  {l l c c c c}
&  & $m_0$ & $\sigma_0$ & $m_1$ & $\sigma_1$ \\
\hline
\textbf{DQKR}&  $\gamma = 0.0$ &  $1.0$ & $0.1124$ & $1.3998$ & $0.0904$ \\
\textbf{RMT} & $\alpha = 0.0$  & $1.0$ & $0.1103$ & $1.3986$ & $0.0869$ \\
\hline
\textbf{DQKR} & $\gamma = 8/N^{3/2}$  & $1.0$ & $0.1004$ & $1.3826$ & $0.0805$\\
\textbf{RMT} & $\alpha = 0.9/N^{1/2}$  & $1.0$ & $0.1000$ & $1.3842$ & $0.0791$ \\
\hline
\textbf{DQKR} & $\gamma = 11/N^{3/2}$ &  $1.0$ & $0.0956$ & $1.3750$ & $0.0753$ \\
\textbf{RMT} & $\alpha = 1.2/N^{1/2}$ & $1.0$ & $0.09578$ & $1.3771$ & $0.0746$ \\
\hline
\textbf{DQKR} & $\gamma = 13/N^{3/2}$ &  $1.0$ & $0.0940$ & $1.3722$ & $0.0731$ \\
\textbf{RMT} & $\alpha = 1.5/N^{1/2}$  & $1.0$ & $0.0929$ & $1.3727$ & $0.0713$ \\
\hline
\textbf{DQKR} & $\gamma  = 0.7$    & $1.0$ & $0.0905$ & $1.3693$ & $0.0682$ \\
\textbf{RMT} & $\alpha = 1.0$  & $1.0$ & $0.0881$ & $1.3682$ & $0.0647$
\end{tabular}
\end{ruledtabular}
\end{table}

 \section{Conclusion} \label{sec-conclusion}
 We have studied ensemble of complex symmetric matrices viz. symm-Ginibre ensemble. The spacing distribution for symm-Ginibre is different from that of Ginibre ensemble. We have also studied the quantum kicked rotor with time reversal invariance in dissipative regime and show that the spacing distribution is same as that of symm-Ginibre ensemble. Thus symm-Ginibre matrices are useful to investigate the universal features and model the dissipative systems with time reversal invariance. We have also introduced the complex matrices that are useful in the study of dissipative system with TRI weakly broken.
 
\bibliography{references.bib}

\end{document}